\documentclass[aps,pra,twocolumn,groupedaddress,showkeys,showpacs,floatfix]{revtex4}
\usepackage{amsmath,amssymb,amsfonts,graphicx,bm,dcolumn}
\begin{document}

\title{A Cascade Model for Particle Concentration 
and Enstrophy in Fully Developed Turbulence with Mass Loading Feedback}
\author{R. C. Hogan}  
\affiliation{Bay Area Environmental Research Institute; MS 245-3 Moffett Field, CA  94035-1000 }  
\email{hogan@cosmic.arc.nasa.gov}   
\author{J. N. Cuzzi}  
\affiliation{NASA Ames Research Center; MS 245-3 Moffett Field, CA  94035-1000}  
\email{jcuzzi@mail.arc.nasa.gov} 
\date{\today}

\begin{abstract}
A cascade model is described based on multiplier distributions determined from
3D direct numerical simulations (DNS) of turbulent particle laden flows, which
include two-way coupling between the phases at global mass loadings equal to
unity. The governing Eulerian equations are solved using psuedo-spectral
methods on up to $512^3$ computional grid points. DNS results for particle
concentration and enstrophy at Taylor microscale Reynolds numbers in the range
34 - 170 were used to directly determine multiplier distributions on spatial
scales 3 times the Kolmogorov length scale. The multiplier probability
distribution functions (PDFs) are well characterized by the $\beta$
distribution function. The width of the PDFs, which is a measure of
intermittency, decreases with increasing mass loading within the local region
where the multipliers are measured. The functional form of this dependence is
not sensitive to Reynolds numbers in the range considered. A partition correlation
probability is included in the cascade model to account for the observed
spatial anticorrelation between particle concentration and enstrophy. Joint
probability distribution functions of concentration and enstrophy generated
using the cascade model are shown to be in excellent agreement with those
derived directly from our 3D simulations. Probabilities predicted by the
cascade model are presented at Reynolds numbers well beyond what is achievable by direct
simulation. These results clearly indicate that particle mass
loading significantly reduces the probabilities of high particle concentration
and enstrophy relative to those resulting from unloaded runs. Particle mass
density appears to reach a limit at around 100 times the gas density.
This approach has promise for significant computational savings in certain applications.
\end{abstract}


\maketitle

\section{Introduction}

The study of turbulent flows incorporating heavy particles in suspension (particles 
with finite stopping times) is an
important endeavor that has both fundamental and practical relevance to many
scientific and engineering problems. Such flows have been investigated mainly
in numerical simulations where detailed statistical analysis of the flow
fields is possible \cite{Squires:4,Squires:1,Squires:2,Kulick:1}   
These
simulations, limited to relatively low Taylor microscale Reynolds numbers
$Re_\lambda$ ($\sim 40$), demonstrated that particles whose fluid response
times are comparable to the lifetime of the smallest turbulent eddies produce a
highly nonuniform field with intense regions of concentration. Preliminary
indications were that the feedback from such concentrations of particles could
locally damp turbulence - however, the role of this ``mass loading" effect in
determining the statistical distributions of particle density and various fluid
scalars has not been thoroughly studied. Experimental investigations of
turbulence modification by particles have demonstrated that the degree of
turbulence  damping increases with particle mass loading and concentration 
\cite{Kulick:1}.

The phenomenon known as intermittency can be described as intense fluctuations,
on small spatial and temporal scales in the turbulent field, that contribute to
the exponential tails of probability distribution functions (PDFs) of scalars
such as velocity increments and gradients \cite{Castaing:1,Dinavahi:1,Kailasnath:1}, 
dissipation \cite{Vincent:1}, pressure \cite{Pumir:1,Lamballais:1}, enstrophy \cite{Jimenez:2,He:1} 
and velocity circulation \cite{Cao:1}. Intermittency in the
density field of preferentially concentrated particles has also been observed
and studied \cite{Hogan:1,Balkovsky:1}.

Although intermittency in turbulence still lacks a complete theoretical
understanding, progress has been made with phenomenological models that capture
intermittency in a cascade process. Richardson \cite{Richardson:1} and 
later Kolmogorov \cite{Komolgorov:1}
suggested that such models might be used to explain the process of eddy
fragmentation initiated by unstable large scale structures in a turbulent fluid.
Intermittency in the context of fragmentation though a cascading process has
been studied for large-scale gravitating masses \cite{Chiueh:1} and velocity
increments in turbulence \cite{Gorokhovski:1}. Simple cascade models were
explored by Meneveau and Sreenivasan \cite{Meneveau:3} and were reviewed by 
Sreenivasan and Stolovitzky \cite{Sreenivasan:1}
The scale similarity of random fields was explored by Novikov \cite{Novikov:2,Novikov:1}, with a focus 
on the energy dissipation cascade. In Novikov's work, the ratio of 
dissipation averaged over two spheres, one embedded within
the other, served as a measure of enstrophy partitioning between larger and smaller
scales. The probability distribution of these ratios, known as multipliers or
breakdown coefficients, was shown to relate to multifractal and statistical
measures (moments) of the velocity and dissipation fields. A recent review of
intermittency in multiplicative cascades  stresses that this theory is a
kinematic description and its connection with the real dynamics remains unclear \cite{Jimenez:1}.

Our previous numerical study of particle concentration in turbulent flows showed
that the particle density field is a multifractal on scales comparable to the
Kolmogorov length scale \cite{Hogan:1}. This result suggests that a deeper
description of the statistical properties of the particle concentration field,
based on multiplier PDFs, may also be possible. Analytical efforts have
suggested that dissipation and vorticity in the fluid phase should be locally
linked with particle concentration \cite{Maxey:1}. Numerical work in this regard
has demonstrated that preferential concentration is statistically
anticorrelated with low vorticity: particles tend to concentrate in regions
where enstrophy is relatively weak \cite{Squires:3,Ahmed:2}.

In this paper we present a cascade model in the spirit of Novikov \cite{Novikov:2,Novikov:1}
that follows the partitioning of positive definite scalars associated with both
the fluid and the particles. Multipliers controlling the partitioning of
enstrophy and particle density at each step in the cascade are drawn from
probability distribution functions (PDFs) which are determined empirically from
direct numerical simulations (DNS). Moreover, the multiplier PDFs are {\it
dependent} on, or conditioned by, the particle mass density or mass loading.
The cascade model then generates joint PDFs for particle concentration and
enstrophy at arbitrary cascade levels. A {\it partitioning correlation
probability} is also applied at each cascade level to account for the observed
spatial anticorrelation between enstrophy and particle concentration \cite{Squires:3,Eaton:1}.

In Section II we describe the cascade model and its parameters, which are
empirically determined from DNS calculations. Details of the DNS equations, and
our numerical methods, are discussed in the Appendix. Results are shown in section
III, including comparisons of joint PDFs of enstrophy and particle
concentration as predicted by the cascade model with those obtained directly
from the DNS results. Cascade model PDF predictions at Reynolds numbers well
beyond the DNS values are also presented. In section IV, we summarize our
results and discuss their implications.
 
\section{ Cascade Model }

 A turbulent cascade can be envisioned as an hierarchical breakdown  of larger
eddies into smaller ones that halts when the  fluid viscosity alone can
dissipate eddy kinetic energy. Eddies or similar turbulent structures such as
vortex tubes are
bundles of energy containing vorticity and dissipation. These
structures start with a size comparable to the integral scale $\Lambda$ of
the flow, and break down in steps to a size  comparable to the Kolmogorov scale
$\eta$ before being dissipated away by viscosity.  The fluid vorticity and
dissipation exhibit spatial fluctuations that increase in intensity as the
spatial scale decreases. This phenonemon is known as intermittency and has been
observed in a variety of processes with strong nonlinear interactions.

In previous numerical and experimental studies, locally averaged intermittent
dissipation fields with scale at or near $\eta$ were used to quantify the
statistical properties of multiplier distributions \cite{Sreenivasan:1}.
Multipliers are random variables that govern the partitioning of a  positive
definite scalar as turbulent structures break down along the cascade. In these
studies the statistical distribution of multipliers (their PDF) were shown to be
invariant over spatial scales that fall within the turbulent inertial range.
Multifractal properties of the cascading field are derivable from such 
multiplier distributions \cite{Novikov:1}, and cascade models based on the
iterative application of multipliers to a cascading variable have been shown to
mimic intermittency.

While invariant with level in the inertial range of a cascade, multiplier PDFs might depend on local
properties of the environment. For instance, Sreenivasan and Stolovitzky \cite{Sreenivasan:1} showed
that  the degree of intermittency in dissipation increases with the degree of
local strain rate, and constructed multiplier distributions for local energy
dissipation {\it conditioned on} the local strain rate. The physical mechanism
behind this effect is believed to be related to vortex stretching dynamics
creating intense bursts of dissipation. 

All the multiplier PDFs measured by Sreenivasan and Stolovitzky \cite{Sreenivasan:1}, whether 
conditioned or unconditioned by local properties, are well characterized by the
$\beta$ distribution function,
\begin{equation}
   \label{beta} p(m) = \frac{\Gamma(2\beta)}{\Gamma(\beta)^2} m^{\beta-1}(1-m)^{\beta-1} 
\end{equation}
where $m$ is the multiplier variable and $\beta$ is a shape controlling
parameter. A large $\beta$ produces a narrow, delta-function-like curve
centered at $m$ = 0.5, whereas $\beta$ = 1 produces a flat distribution between
 $m$ = 0 and 1. These limits for $\beta$ correspond to uniform and highly
intermittent processes respectively. In conditioned multipliers, the value
of $\beta$ varies with some local property of the fluid.
 
Concentration of particles in turbulence is a result of  the active dynamics of
eddies on all scales. The  process depends on the scale of the eddies and the
corresponding  particle response to those eddies. Intense particle density
fluctuations,  akin to intermittency, were observed in a previous numerical
study  where it was also shown that nonuniform particle concentrations have
multifractal scaling properties \cite{Hogan:1}. These results strongly
suggest that a phenomenological cascade model based on multipliers may
adequately describe the particle density field. Simulations that have  included
particle feedback on the fluid through the mass loading  effect show that
damping of local turbulence occurs \cite{Squires:1,Elghobashi:2}. 
The latter have shown that vorticity dynamics is affected 
locally by particle feedback. This interplay between the phases could attenuate 
vortex stretching and, thereby, diminish local turbulent intermittency. 
Multiplier distributions conditioned on local mass loading should therefore be an 
integral part of a realistic fluid-particle cascade model.

\subsection { Two-Phase Cascade model }

 Below we describe a two-phase cascade model that incorporates  simultaneous
multiplier processes for particle concentration $C$ and fluid enstrophy $S$, in
addition to a process that models their spatial anticorrelation.  The
multiplier distributions are conditioned by the local particle concentration,
as determined empirically from DNS fields  equilibrated to  $Re_\lambda$ = 34,
60, 107, and 170. The spatial  anticorrelation was also quantified from these
fields. Local measures of particle concentration ($C$) and enstrophy ($S$) used
are defined in the Appendix.

\begin{figure}[h]
\begin{center}
\includegraphics[width=3.2in]{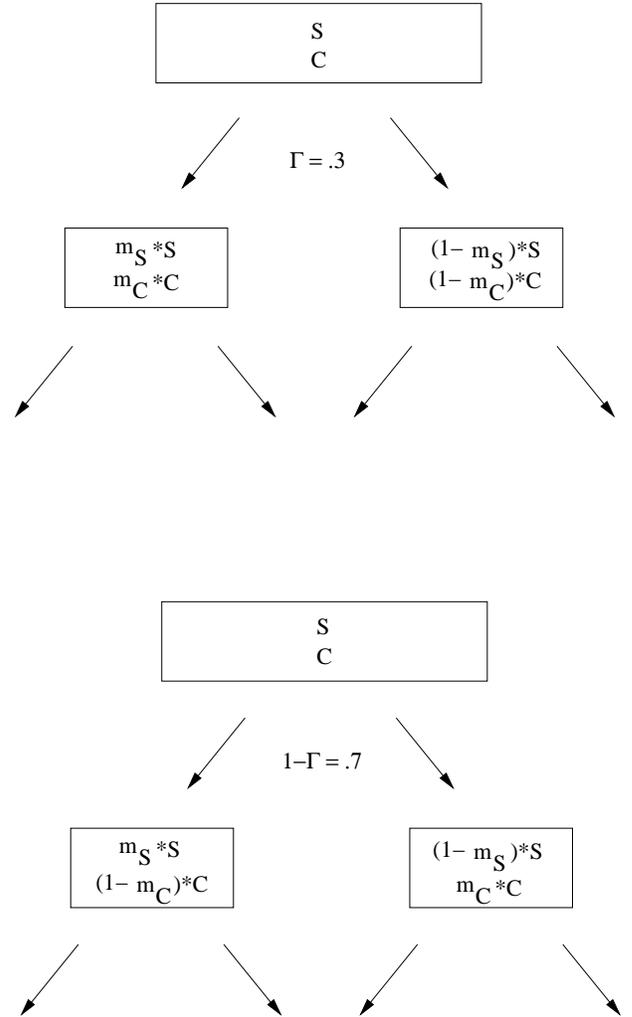}
\caption{\label{fig1}Figure depicting the breakdown of a parcel of enstrophy ($S$) and     
particle concentration ($C$) into two parcels each with half the volume of     
the parent.      The corresponding multipliers $m_{S}$ and $m_{C}$ are
assumed to be greater than 0.5 in this figure. These measures are broken down
and distributed      between the two parcels in one of two ways - the larger
portions are partitioned together with probability $\Gamma$= 0.3      (upper
figure), or in opposite directions with probability $1-\Gamma$= 0.7 (lower
figure). }
\end{center}
\end{figure}

A schematic illustration of our two-phase partitioning process is shown in 
FIG.~\ref{fig1}. The cascading vector $(S,C)$ has components representing enstrophy
and  particle concentration. Initially the components are assigned the value
unity  and are associated with a common cell having a volume of unity. Each
component   is partitioned into two parts;  $(m_{S}S, (1-m_{S})S)$ and
$(m_{C}C, (1-m_{C})C)$, respectively, where $m_{S}, m_{C}$ are multipliers for $S$ and $C$
whose values are   between zero and one inclusive and are random members of 
 the corresponding multiplier distributions. The parts are associated with two
daughter cells each containing half the volume of the starting cell.   In the example
shown in FIG.~\ref{fig1},
$m_{S}$ and $m_{C}$ are assumed to be greater than 0.5.   The largest parts of
$S$ and $C$ are placed in the same daughter cell with probability $\Gamma$  (and in
different cells with probability $1-\Gamma$).   This partitioning process is
repeated for each daughter cell down the cascade until the ratio of the
daughter cell size to the   initial cell size equals a specified cutoff. When
this cutoff is set to the   ratio of the turbulent lengthscales $\Lambda$ and 
$\eta$, the   cascade corresponds to turbulence characterized by
$Re_{\lambda} \sim (\Lambda / \eta)^{2/3}$ \cite{Frisch:1}.


\subsection{ Conditioned Multipliers }

The parameters of the cascade model are empirically derived from    the
particle density and enstrophy fields $C$ and $S$ as calculated by DNS (see Appendix). 
The simulation parameters for four DNS runs representing   
$Re_\lambda$ =  36, 60, 104, and 170 are shown in Table ~\ref{tab1}.
The turbulence kinetic energy $q$, the volume averaged dissipation $\epsilon$, and $\Lambda$ 
are calculated from the 3-D turbulent energy spectrum $E(k)$ and kinematic viscosity $\nu$,

\begin{equation}
q = \int_{0}^{\infty}E(k)dk  
\end{equation}

\begin{equation}
\epsilon = 2\nu\int_{0}^{\infty}E(k)k^2dk 
\end{equation}

\begin{equation}
\Lambda = \frac{3\pi}{4q}\int_{0}^{\infty}\frac{E(k)}{k}dk  
\end{equation}

where $k$ is wavenumber.
$k_{max} = \frac{\sqrt{2}}{3}$ times the number of computational nodes per side is the 
maximum effective wavenumber. Thus $k_{max}\eta > 1$ indicates an adequate resolution of
the Kolmogorov scale.


\begin{table}[h]
\begin{center}
\begin{tabular}{|c|c|c|c|c|}
\hline
Parameter & Case I & Case II & Case III & Case IV \\
\hline
Nodes/side            & 64       & 128      & 256       &   512  \\
$\nu$                 & .01      & .003     & .0007     &  .0002 \\
$Re_{\lambda}$        & 34       & 60.      & 104       &   170  \\
$q$                   & 1.5      & .65      & .28       &   .14  \\
$\frac{\epsilon}{\nu}$ & 23.      & 22.8  & 22.4   &   23   \\
$ k_{max}\eta$         & 1.4      & 1.5      & 1.45      &   1.56  \\
$\frac{\Lambda}{\eta}$ & 14.1 & 23.3 & 45.8 & 86.2 \\
$\Gamma$                & .31      & .29      & .27       &  .32   \\ 
$D$                   &  .0001   & .00003   & .000007   &  .000002 \\ 
$\nu_{p}$             &  .001    & .0003.   & .00007    &  .00002  \\
\hline
\end{tabular}
\caption{\label{tab1}Case Parameters for DNS runs. The quantities $D$ and $\nu_p$
are defined in the Appendix. Other quantities above are defined in Section II.}


\end{center}
\end{table}

\begin{figure}[h]
\begin{center}
\includegraphics[width=3.2in]{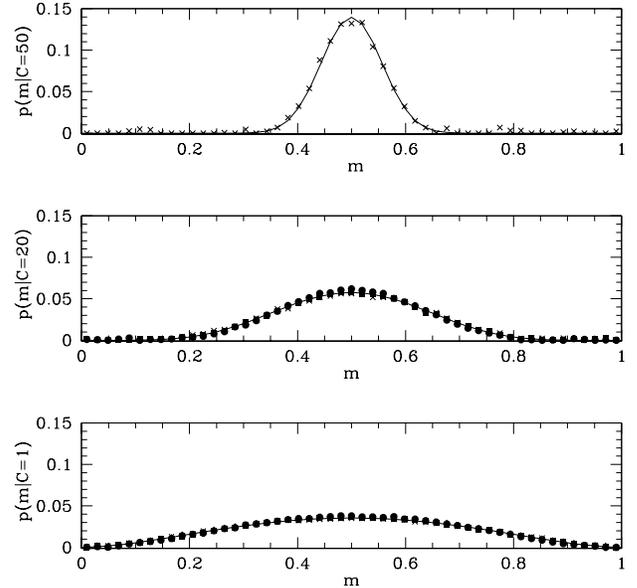}
\caption{\label{fig2} Empirically determined conditional multiplier distributions $p(m|C)$ for  
   particle concentration at three different      mass loading values, $C$ = 1,
20 and 50. The distributions are obtained from bifurcations of      cells with
a spatial scale equal      to $3\eta$.      Results at $Re_{\lambda}$  = 34 (
square ), 60 (triangle),      107 (circle) and 170 ( cross ) are overlain. Only
the simulation with $Re_{\lambda} = 170$ provided results for  $C$ = 50. At
each mass loading      the $p(m)$ at all Reynolds numbers are very well
approximated      with the $\beta$ distribution function ( solid line ).     
The distribution widths narrow as the mass loading increases,      indicating a
decrease in the intermittency.
}
\end{center}
\end{figure}

\begin{figure}[h]
\begin{center}
\includegraphics[width=3.2in]{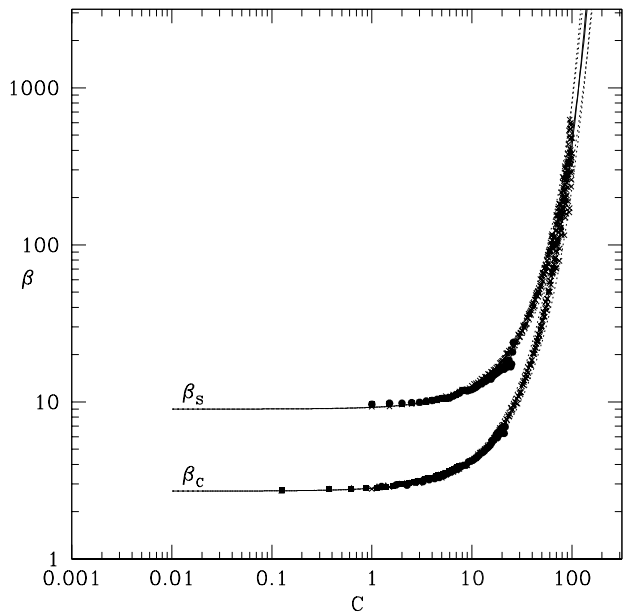}
\caption{\label{fig3}The $\beta$ parameters as functions of local mass loading $C$ for enstrophy  
and particle concentration at $3\eta$.  Results for all      DNS cases are
indicated as described in FIG.~\ref{fig2}.      A least squares      fit of an
exponential function to the points over the      entire mass loading range is
shown ( solid line ). Dashed lines correspond to the upper      and lower limits
of the function, and are derived using the $2\sigma$ errors of $p_2$ and $p_3$.
}
\end{center}
\end{figure}

 The 3-D DNS computational box  is uniformly subdivided into spatial cells
$3\eta$ on a side, and the average  value of $C$ and $S$ is determined for each
cell ( see Appendix ).  The cells are divided into groups associated with disjoint ranges of
$C$.  Each cell is then divided into two parts of equal volume  
and averages for $C$ and $S$ are determined for each
part.  The $C$ and $S$ multipliers for each cell are evaluated as the ratio of
these averages  to the averages in the parent cell.  
A conditional multiplier
distribution $p(m)$ is then determined for each binned value of $C$ from the
corresponding set of cell multipliers.  Plots of $p(m)$ for three values of $C$
are shown in FIG.~\ref{fig2}. The points  represent distributions derived from all DNS
runs and the solid lines are least squares  fits to the $\beta$ distribution
function (Eq.~\ref{beta}). For the lower values of $C$,
$Re_\lambda$-independence is apparent; only the $Re_\lambda = 170$ case
provided data for the largest $C$ range. The plots clearly indicate that  the
intermittency in $C$ is reduced (multiplier PDFs narrow) as $C$ is increased.
Derived values of $\beta_C(C)$ and $\beta_S(C)$ are shown as a function of $C$
in FIG.~\ref{fig3}.  Least squares fits to the functional form $p_1\exp(p_2C^{p_3})$
are drawn as solid lines and the best fit parameter values for this function
are tabulated in Table ~\ref{tab2}. Bounding curves (dashed lines) are defined by
setting $p_2$ and $p_3$ to their $2\sigma$ limits, to establish a plausible
range of uncertainty in the predictions. 


\begin{table}[h]
\begin{center}
\begin{tabular}{|c|c|c|c|}
\hline
$Scalar$ & $p_1$ & $p_2$ & $p_3$ \\
\hline
$C$   & 2.7   & .045  & 1.02   \\   
$S$   & 9.    & .03   & 1.06   \\
\hline
\end{tabular}
\caption{\label{tab2} $\beta$ model parameters } 
\end{center}
\end{table}

It is certainly of interest that such large solid/gas mass loadings as $C=100$
appear in the DNS runs at all, given published reports that particle mass
loading significantly dampens turbulent intensity even for mass loadings on the
order of unity \cite{Squires:4,Kulick:1}. These diverse results might be
reconciled since the particles we study herein are all far smaller than the
Kolmogorov scale
and also have only a very small lag velocity relative to
the gas. Recall that we force the turbulence, as might be the case if it were
being constantly forced by energetic sources operating on
larger scales than our computational volume. 
However, FIG. ~\ref{fig3} strongly suggests an upper limit for $C$ ( $\sim 100$ ) for
both $\beta_S$ and $\beta_C$. 


The cascade anticorrelation parameter $\Gamma$ was determined by counting the
number of   parent cells within which the larger partitions of $C$ and $S$ were
found to share  the same daughter cell. This number divided by the total number
of parent cells defines $\Gamma$.   The derived $\Gamma$ value is approximately
constant across the DNS cases, as indicated in Table ~\ref{tab1}. Operationally, the
$\Gamma$ used in the cascade model was determined by taking a simple average of
the $\Gamma$ values in Table ~\ref{tab1}.

  Overall, the invariance of $\Gamma$ and the $\beta_C(C)$ and $\beta_S(C)$ 
functions across our range of $Re_\lambda$ justifies their treatment as level 
independent parameters in the two-phase cascade model. 
One caveat remains, which would be of interest to address in future work. While
it {\it has} been shown that multiplier distributions leading to $\beta_C$ and
$\beta_S$ are level-invariant over a range of scales within an inertial range
~\cite{Sreenivasan:1}, our simulations were numerically restricted to values of $Re$ in which
the inertial range has not yet become fully developed. Our reliance on the
smallest available scales of 3$\eta$ to 1.5$\eta$ (those providing the largest
available intermittency) might lead to some concern that they were already
sampling the {\it dissipation} range of our calculations, and thus may not be
appropriate for a cascade code. We tested this possibility by calculating
multipliers for the next largest level bifurcation (6$\eta$ to 3$\eta$) for the
$Re_{\lambda}=170$ case. The $\beta$ values for those multiplier distributions
are slightly larger in value, but consistent with the $C$-dependence shown in
FIG. ~\ref{fig2} (6$\eta$ scales don't provide good distribution functions beyond $C
\sim$ 15). Thus we believe that for the purpose of demonstrating this
technique, and for the purpose of estimating the occurrence statistics of $C$
under particle mass loading, our results are satisfactory. For applications
requiring quantitatively detailed and/or more accurate $P(S,C)$, it would
certainly be of interest to extend the DNS calculations to larger $Re$, at
which a true inertial range might be found.

\section { Model Results }

The 2D joint probability distribution function or PDF of concentration and
enstrophy, a fractional volume measure, was generated from the cascade model
and compared with results derived
directly from numerical DNS simulations. The basic probability density
$P(S,C)$ gives the fractional volume occupied by cells having enstrophy $S$ and
concentration $C$, per unit $S$ and $C$; thus the fractional volume having $C$
and $S$ in some range $\Delta S, \Delta C$ is $P(S,C) \Delta S \Delta C$.
For quantities varying over orders of magnitude, it is convenient to adopt
$\Delta S = S$ and $\Delta C = C$, and we will present the results in the
form $P(S,C)SC$.
                                                                        
We started by binning results at spatial scale $3\eta$, obtained from
the semi-final level of a cascade model run, into a uniform logarithmic grid of
$S,C$ bins each having width $\Delta(logS) = \Delta(logC) = \delta$, with 
corresponding values of $\Delta S$ and $\Delta C$. 
The number of $3\eta$ cells accumulated in each bin was normalized
by the total number of such cells in the sample to convert it to a
fractional volume $\Delta V(S,C) = P(S,C)\Delta S \Delta C$. Then
\begin{equation}
\frac{\Delta V(S,C)}{\delta^2} =
\frac{ P(S,C)\Delta S \Delta C}{\Delta (logS) \Delta (logC)}
\rightarrow P(S,C)SC \hspace{0.1 in} {\rm as} \hspace{0.1 in} \delta \rightarrow 0.
\end{equation}
In practice of course, the binning ranges $\delta$ are not vanishingly small.
 
\begin{figure}[h]
\begin{center}
\includegraphics[width=3.2in]{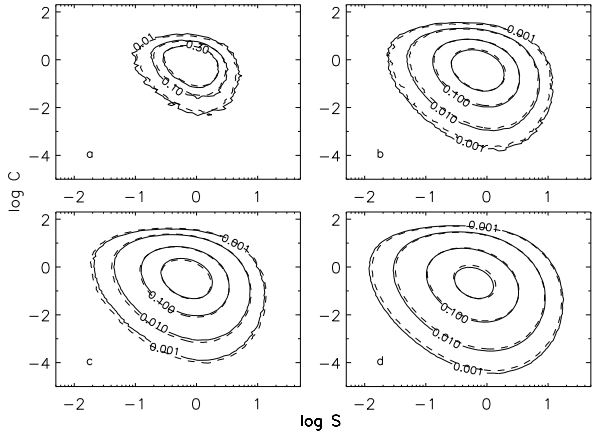}
\caption{\label{fig4}Comparisons of cascade model predictions of $P(S,C)SC$ with DNS     
results at $Re_\lambda$ = 34 (a), 60 (b), 107 (c) , and 170 (d).     
Contours indicate probabilities  .001, .01, .1 and .3. Dashed contours are
cascade model predictions and solid ones are DNS results.}
\end{center}
\end{figure}

The plots in FIGs. ~\ref{fig4} ~\ref{fig5} and ~\ref{fig7} then, show the PDF as the 
volume fraction $P(S,C)SC$. Cascade
levels 9, 12, 15, and 18    correspond approximately to the $Re_\lambda$ of the
four simulation cases shown in Table ~\ref{tab1}. These levels were determined from the
ratio of $\Lambda$ and $\eta$ for each case: level = $3 {\rm log}_2(\Lambda/
\eta)$. The factor $3$ accounts for cascade bifurcations of 3D cells, because
it takes three partitionings, along three orthogonal planes, to generate eight
subvolumes of linear dimension one-half that of the parent volume. That is,
$2^{level}$ is equal to the number of $\eta$ cells within a 3D volume having
linear dimension $\Lambda$ and $(2^{level/3})^{2/3}$ is the corresponding $Re_{\lambda}$.
The number of cascade realizations is, in turn,
equal to the product of the number of $\Lambda$-size volumes in the
computational box and the number of simulation snapshots processed. In general
it is difficult to generate DNS results with a    ratio of $\Lambda$ and $\eta$
that is an exact power of two.    In order to correctly compare DNS simulations
with the cascade model it was    necessary to interpolate between two cascade
generated $P(S,C)SC$ computed    at scale ratios (levels) that bracketed the
ratios that were    actually simulated. In FIG.~\ref{fig4} we compare iso-probability  contours
of $P(S,C)SC$    predicted by cascade models representing the four DNS cases
with the same contours derived directly from the simulated $S$
and $C$ fields. The agreement is very good.


\subsection { Predictions at higher Reynolds number }

The cascade model was used to generate PDFs at deeper levels in order to
assess the effect of mass loading on the probabilities of high $C$ and $S$. We
generated 256 realizations of a level 24 cascade, 20 realizations of a level 30
cascade, and one realization of a level 36 cascade.

\begin{figure}[h]
\begin{center}
\includegraphics[width=3.2in]{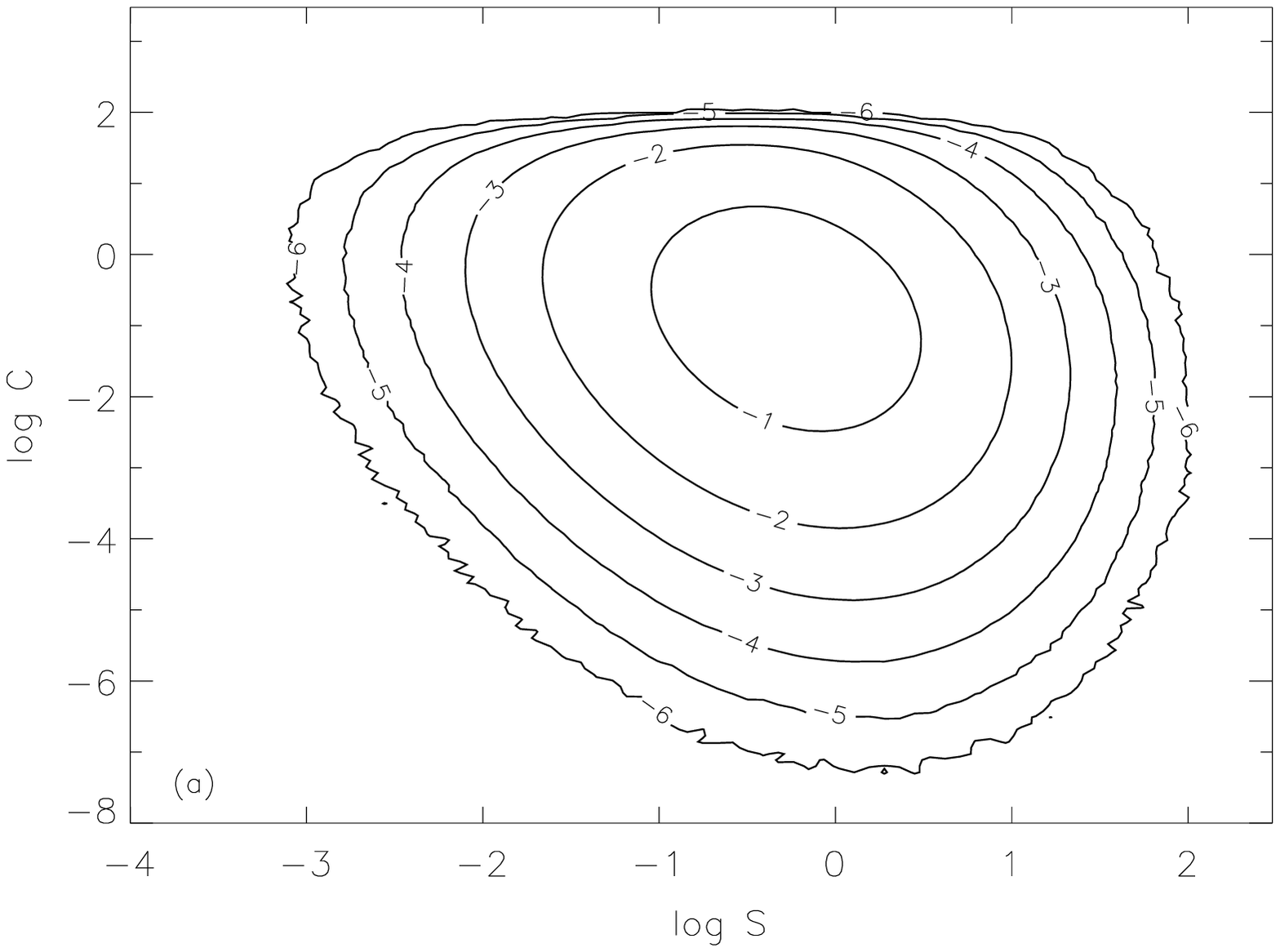}
\includegraphics[width=3.2in]{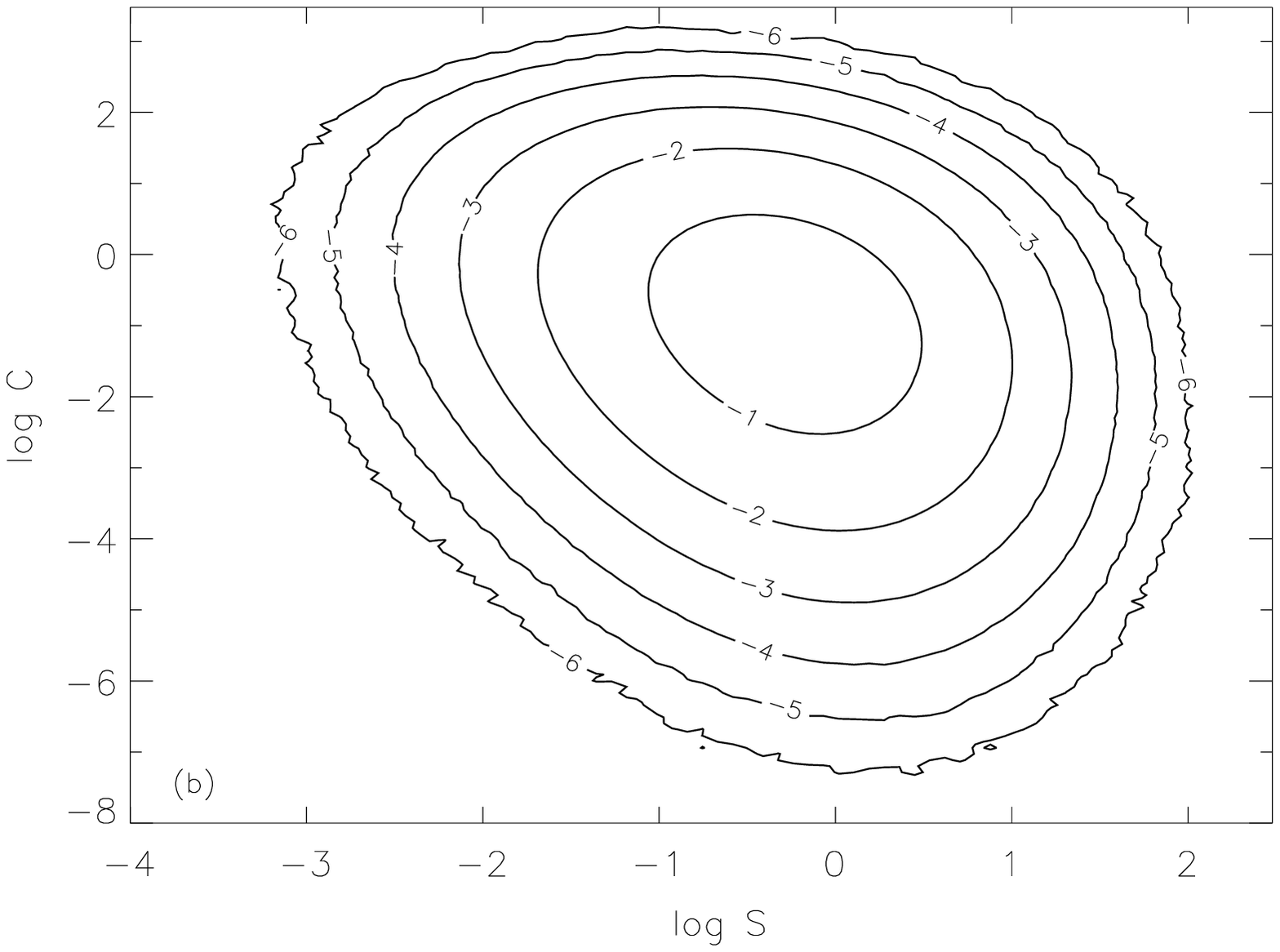}
\caption{\label{fig5}(a) Cascade model predictions for a 24 level case, taken to lower
probability levels, using 256 realizations of the cascade. Contours are labeled
by $log(P(S,C)SC)$. Note the crowding of contours at high $C$
values, indicating the high-$C$ limit of the process under conditions of
mass loading.(b) A control cascade to level 24, as in FIG.~\ref{fig5}(a), with
conditioning turned off. The difference between (a) and (b) clearly shows the
``choking" effects of particle mass loading on intermittency in $C$. }
\end{center}
\end{figure}

FIG.~\ref{fig5}(a) shows the average of 256 realizations of a 24 level cascade, taken to
lower probability values. 
The pronounced crowding of the contours at the top of
the figure indicates the effect of particle mass loading on reducing the
intermittency of $C$ at high values of $C$. For comparison, FIG.~\ref{fig5}(b) shows a
control run of a 24 level cascade with all conditioning turned off. In this
control case, the exponential tails characterizing intermittent fluctuations
are seen at both low {\it and} high $C$.


\begin{figure}[h]
\begin{center}
\includegraphics[width=3.2in]{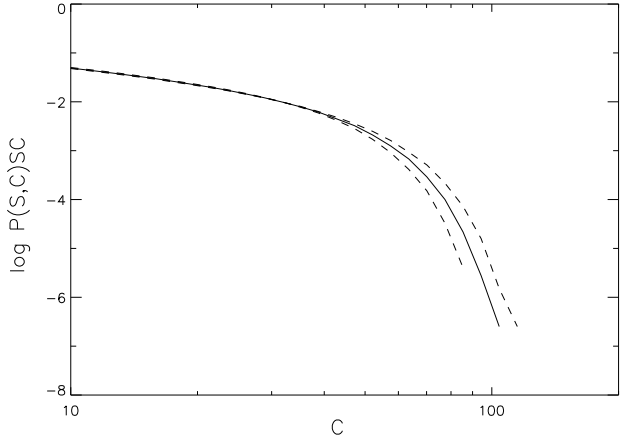}
\caption{\label{fig6}1D cuts through 
the mode of the PDF of FIG.~\ref{fig5}(a) parallel to the $C$ axis, showing the effects 
of uncertainty in the conditioning curve $\beta_C(C)$. The solid curve is the 
nominal model and the dashed curves are obtained by allowing the parameters
$p_2$ and $p_3$ to take their $2\sigma$ extreme values.}
\end{center}
\end{figure}

In order to evaluate the effect of the uncertainties in the extrapolations of
the $\beta$ curves for $C$ and $S$ on the PDF, two cascade runs to level 24
were generated using the parameters for the upper and lower dotted curves in FIG.~\ref{fig3}. 
In FIG.~\ref{fig6} we show cross-sections of the PDFs produced by these runs
along the $C$ axis through the distribution modes to compare with the same
cross-section for a run using the nominal parameters in Table ~\ref{tab2}. 
Both models
diverge from the mean model beyond $C > 40$, with the upper (lower) curve
corresponding to the outside (inside) $\beta_C(C)$ and $\beta_S(C)$ bounds in
FIG.~\ref{fig3}. Figure~\ref{fig6} indicates that the sensitivity of the PDF to the $\beta$
model parameters at the $2\sigma$ level is only apparent at large $C$, and all
models show a sharp dropoff in the probability for $C >$ 100.


\begin{figure}[h]
\begin{center}
\includegraphics[width=3.4in]{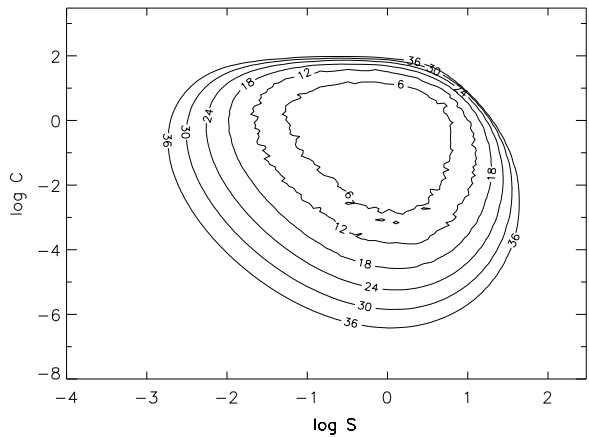}
\caption{\label{fig7}Cascade model predictions for $P(S,C)SC = 5 \times 10^{-4}$ for levels
6, 12, 18, 24, 30, and 36. Contour labels indicate the cascade levels.}
\end{center}
\end{figure}

A crowding effect similar to the one seen in FIG.~\ref{fig5}(a) is shown in FIG.~\ref{fig7} for
iso-probability contours equal to $5 \times 10^{-4}$, for cascade levels 6, 12,
18, 24, 30 and 36. 


\begin{figure}[h]
\begin{center}
\includegraphics[width=3.2in]{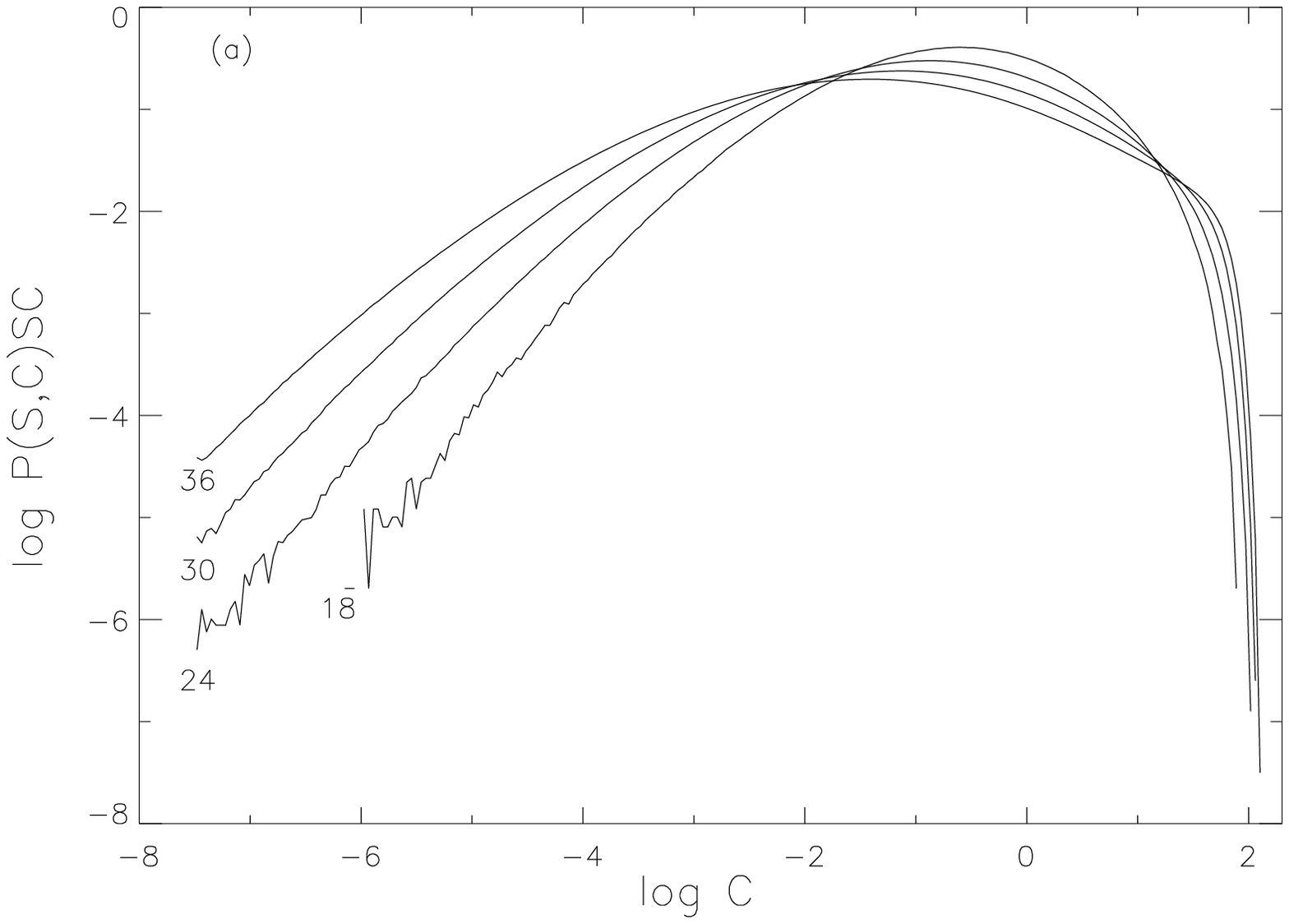}
\includegraphics[width=3.2in]{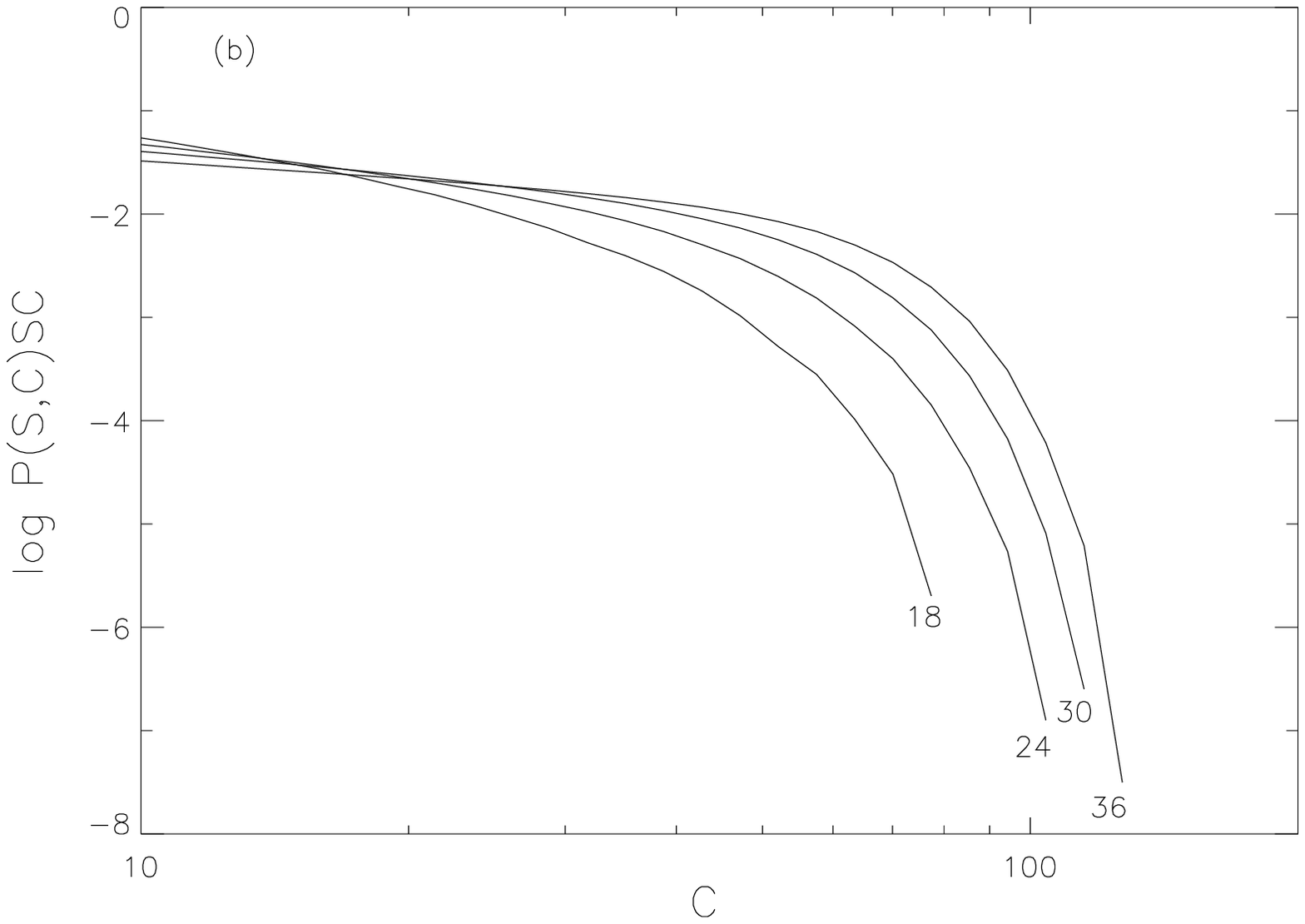}
\caption{\label{fig8}(a) 1D global cuts through the cascade model PDFs $P(S,C)SC$ for 
runs with 18, 24, 30, and 36 levels. (b) closeup of 1-D cuts through high-$C$ regime. } 
\end{center}
\end{figure}

Figures~\ref{fig8}(a) and ~\ref{fig8}(b) compare 1D cuts through the modes of the PDFs for
cascades of 18 - 36 levels, indicating that going to deeper levels (higher
$Re_{\lambda}$) results in larger intermittency at the low-$C$ end (as expected), retaining
the exponential tail characteristic of intermittent processes, but the
highest particle concentration end of the distribution is extended more slowly. Certainly
at the order of magnitude level, a particle mass loading ratio of 100 times the
gas density appears to be as high as preferential concentration can produce.
This result could be inferred directly from inspection of the conditioned
$\beta$ distributions of FIG.~\ref{fig3}.


\section {Summary}

A two-phase cascade model for enstrophy and particle concentration in 3-D,
isotropic, fully developed turbulence with particle loading feedback
has been developed and tested. Multiplier
distributions for enstrophy and particle concentration were empirically
determined from direct numerical simulation fields at Taylor scale Reynolds
numbers between 34 and 170. These simulations included `two-way' coupling
between the phases at global particle/gas mass loadings equal to unity. The
shape of all multiplier distributions is well characterized by the $\beta$
distribution function, with a value of $\beta$ that depends systematically on
the local degree of mass loading. The values of $\beta$ increase monotonically
with mass loading and begin to rapidly increase at mass loadings greater
than $100$.

The $C$-dependent multiplier distributions were used as input to a cascade
model that simulates the breakdown, or cascade, of enstrophy $S$ and particle
concentration $C$ from large to small spatial scales. The spatial
anticorrelation between enstrophy and particle concentration was empirically
determined from 3D DNS models and shown to be constant with $Re_{\lambda}$.
This constant was used as a correlation probability governing the relative
spatial distribution of $S$ and $C$ at each bifurcation step in the cascade
model.

The cascade model we have developed clearly reproduces the statistical
distributions and spatial correlations observed in our DNS calculations.  The
cascade parameter values we have derived appear to be universal within the
range of  $Re_{\lambda}$  of our simulations. 
We thus speculate that they
can be used to predict approximate joint probabilities of enstrophy and particle
concentration at higher Reynolds numbers, at great savings in computer time.
For example, a typical DNS run to $Re_{\lambda} = 170$ takes about 170 cpu
hours on an Origins 3000 machine, while a cascade model to an equivalent level
takes 0.1 cpu hours. 

We have presented joint probabilites of $S$ and $C$ derived
from cascade runs up to level 36. The contours shown in FIG.~\ref{fig5}(a) 
 and FIG. ~\ref{fig6} clearly show
the effects of particle mass loading on the probability distribution
functions of $C$ in the regimes where $C$ is large. It appears that particle
mass loadings greater than 100 are rare in turbulent flows.

The properties of the cascade rest on the physics of our DNS simulations, and we speculate that
two separate effects are involved. First, particle mass loading dampens fluid
motions of all types, decreasing vorticity stretching and all other forms of
ongoing eddy bifurcation which are needed to produce intermittency. Second, as
a byproduct of this, particle mass loading may alter the Kolmogorov timescale
locally and shift the most effectively concentrated particle Stokes number $St$
to a larger value than that characterizing particles already lying in the local
volume, reducing the probability of preferentially concentrating the local
particles any further.

{\it Caveats and Future Work:}

As described in section II, our multiplier distributions were taken from the 
most numerous cells, with the largest intermittency, which are at the smallest 
scales possible (furthest from the forcing scale). At Reynolds numbers 
accessible to DNS, a true inertial range is only beginning to appear, and 
while, sampling at the smallest spatial scales possible, we are 
as closely approaching the asymptotic values within the true inertial range as 
possible, where level-independence has been demonstrated in the past ~\cite{Sreenivasan:1}, it
is possible that our values are subject to inaccuracy by virtue of being 
sampled too close to the dissipation scale. Any such inaccuracy will affect our 
cascade results quantitatively but not qualitatively. As computer power 
increases, it would be a sensible thing to continue experiments like these at 
higher $Re_{\lambda}$. 

A more general model that treats enstrophy and strain as independent cascading 
scalars {\it might} allow for a higher-fidelity particle concentration cascade, 
since $C$ is known to be linked to the difference between these two scalars 
~\cite{Maxey:1} (the so-called second invariant tensor {\it II}). However, such an effort 
would introduce further complexity of its own, as {\it II} is no longer 
positive definite. We consider the development of such a model a suitable task 
for future work.

      
\appendix*
\section{} 
   We used an Eulerian scheme developed by Dr. Alan Wray    
to solve the coupled set of    fluid/particle equations used in this study.
This was done to    maximize the computational efficiency of the calculations
and, more    importantly, to accurately evaluate multipliers over the wide   
range of particle concentrations and enstrophies expected.    In this study the
effects of particle collisions and external forces on the    particles (e.g., gravity)
are not considered. The turbulence is spectrally forced at $k = \sqrt{14}$ such that 
moments of the Fourier coefficients of the force satisfy isotropy up to the fourth
order.  
The instantaneous Navier-Stokes equations describing the conservation of
mass and    momentum for an incompressible fluid are

\begin{equation}
  \mathbf{\nabla \cdot U} = 0  \label{a1}
\end{equation}

\begin{equation}
  \frac{\partial{\mathbf{U}}}{\partial{t}} + \mathbf{(U \cdot \nabla ) U} = \
   - \frac{\mathbf{\nabla} P }{\rho_f} + \nu \mathbf{\nabla^2 U} - \alpha 
\frac{\rho_p}{\rho_f} 
   \mathbf{(U - V)} \label{a2}
\end{equation}
where $\mathbf{U}$ is fluid velocity, $\mathbf{V}$ is particle velocity, $\rho_f$ and $\rho_p$ 
are the fluid and
particle mass densities, $\nu$ is fluid viscosity, $P$ is pressure, and
$\alpha$ is the inverse of the particle gas drag stopping time $\tau_p$.

 The compressible equations for the particles are 

\begin{equation}
  \frac{\partial{\rho_p}}{\partial{t}} + \mathbf{\nabla}(\rho_p\mathbf{V}) = \
  D\mathbf{\nabla^2 \rho_p} \label{a3}
\end{equation}

\begin{equation}
  \frac{\partial{(\rho_p \mathbf{V})}}{\partial{t}} + \mathbf{\nabla}(\rho_p 
\mathbf{VV}) = \
  \nu_p\mathbf{\nabla^2}(\rho_p \mathbf{V}) + \alpha \rho_p \mathbf{(U - V)} \label{a4} 
\end{equation}
   where $\nu_p$ is a ``particle 
viscosity", and  $D$ is 
a ``particle diffusivity".
   The particle diffusivity and viscosity terms numerically smooth out particle 
mass and 
   momentum, alleviating the formation of steep gradients of $\rho_p$ that can 
lead to 
   numerical instabilities {\it eg.} \cite{Johansen:1}.

   The right hand sides of Eqs. \ref{a2} and \ref{a4} contain phase coupling terms
   which are linearly dependent on $\mathbf{(U-V)}$.   
   The linear form of the coupling follows from the assumptions that the 
   particle size is much less than $\eta$, and that the material density of the 
particles is much greater than $\rho_f$ \cite{Squires:1}.
   Additional contributions to the particle-gas couplings involving pressure,
viscous and Basset forces \cite{Elghobashi:2} have not
   been added since they are expected to be weak in our size regime of 
interest. 
   The particle field is introduced with a constant mass density and an initial 
velocity given by the local gas velocity in a field of statistically stationary turbulence.
   All runs are continued until the particle statistics (RMS of conentration 
distribution) have equilibrated.

   The particle Stokes number $St$ is defined relative to the Kolmogorov time 
scale $\tau_{\eta}$ as $St = \tau_p / \tau_\eta$, and $\Phi =  M_p/M_f$ is the 
global mass loading, where $M_p$ and $M_f$ are the total mass of particles and 
fluid respectively.
   In this study $\rho_f$, $St$, and $\Phi$ are set to unity, $D/\nu = 0.01 $, 
   and $\nu_p/\nu = 0.1$. Explicitly setting $St = 1$ guarantees that the 
particles
   are preferentially concentrated. When $\Phi$ is unity, 
   $\rho_p$ is a surrogate for the local mass loading or local concentration
   factor $C$.  The values of $\nu_p$ and $D$
   minimize the diluting effects of numerical particle diffusion 
   while preventing numerical blowups; their values were determined 
   from a set of DNS runs in which their values were decreased  systematically  
   until numerical instabilities set in.

   Eqs. \ref{a1} - \ref{a4} are solved using psuedo-spectral methods commonly
   used to solve Naviers-Stokes equations for a turbulent fluid.  The Fast
   Fourier Transform (FFT) algorithm is used to efficiently evaluate
   the dynamical variables $\mathbf{U}$, $\mathbf{V}$ and $\rho_p$ on a 3D 
uniform grid of 
   computional nodes with periodic boundary conditions. 
   The computational 
algorithm is parallelized using MPI and is written in Fortran 90. All runs for 
this study were executed on  SGI Origins supercomputers with up to 1024 
processors. 
   
   Enstrophy is defined as

   \begin{equation}
    S = \frac{1}{2}\sum_{ij}({ \partial_i \mathbf{U_j} - \partial_j
\mathbf{U_i} })^2  \label{a5}
   \end{equation}
where $i,j$ are summed over the three coordinate dimensions of $\mathbf{U}$.

  The local spatial average of a scalar over a sample volume is estimated as,

\begin{equation}
  \frac{1}{n dv}\sum_{i}^{n} F_i dv \label{a6}       
\end{equation}
where $F_i$ is the scalar's value on computational node $i$ centered   within a
cube of volume $dv$ and the sum   is over all $n$ nodes covering the sample
volume.  We normalized this average by the global average value to get a
quantity that measures the scalar's  local value relative to its mean.
In this paper $C$ and $S$ will denote
normalized spatial averages of particle concentration and enstrophy over
cubes $3\eta$ on a side.

\begin{acknowledgments}

We are very grateful to Dr. Alan Wray for providing the 3-D code and for   
useful comments on its use. We thank Robert Last for parallelizing the cascade
code on the SGI Origins 3000. We also would like to thank the consultants and
support staff at the NAS facility for providing invaluable assistance, and the
Science Mission Directorate of NASA for generous grants of computer time. 
We thank Prof. K. Sreenivasan for several helpful conversations in the
preliminary stages of this project and the internal reviewers Drs. Alan Wray 
and Denis Richard for their suggestions
for improving the manuscript. This research has been made possible by
a grant from NASA's Planetary Geology and Geophysics program.

\end{acknowledgments}


\end{document}